\documentclass[journal]{IEEEtai}

\usepackage[colorlinks,urlcolor=blue,linkcolor=blue,citecolor=blue]{hyperref}

\usepackage{color,array}

\usepackage{graphicx}
\usepackage{academicons}
\usepackage{amsmath}
\usepackage{xcolor}
\usepackage{longtable}
\usepackage{caption}

\begin{document}

\title{Generative AI in Modern Education Society}

\author{Sanjay Chakraborty
\thanks{Sanjay Chakraborty is working in the Department of Computer and Information Science (IDA) at Linköping University, Sweden and Department of Computer Science \& Engineering, Techno International New Town, Kolkata, India (Email: schakraborty770@gmail.com).}}


\maketitle

\begin{abstract}
Transitioning from Education 1.0 to Education 5.0, the integration of generative artificial intelligence (GenAI) revolutionizes the learning environment by fostering enhanced human-machine collaboration, enabling personalized, adaptive and experiential learning, and preparing students with the skills and adaptability needed for the future workforce. Our understanding of academic integrity and the scholarship of teaching, learning, and research has been revolutionised by GenAI. Schools and universities around the world are experimenting and exploring the integration of GenAI in their education systems (like, curriculum design, teaching process and assessments, administrative tasks, results generation and so on). The findings of the literature study demonstrate how well GenAI has been incorporated into the global educational system. This study explains the roles of GenAI in the schooling and university education systems with respect to the different stakeholders (students, teachers, researchers etc,). It highlights the current challenges of integrating Generative AI into the education system and outlines future directions for leveraging GenAI to enhance educational practices. 
\end{abstract}

\begin{IEEEImpStatement}
The underutilization of GenAI in smart education today lies in its limited application for fostering creativity, critical thinking, and adaptability, indicating the need for comprehensive research on its potential to transform teaching methods and learning environments. This review is motivated by the need to explore the integration of generative artificial intelligence (GenAI) into the modern smart education system, leveraging cutting-edge tools and technologies. This paper examines key case studies and relevant literature on AI-driven modern education systems, technologies, and sociological impacts. It also identifies several challenges related to the implications of modern AI in education 5.0, including how they affect learning environments, teaching methods, and research \& development. Our findings suggest that education is evolving from simple automation toward genuine collaboration between AI and humans, offering new research opportunities. Future work will focus on exploring targeted applications of generative AI within Education 5.0 frameworks.
\end{IEEEImpStatement}

\begin{IEEEkeywords}
Artificial Intelligence, Education 5.0, Generative AI, Smart Learning, Smart Teaching.
\end{IEEEkeywords}

\section{INTRODUCTION}
\begin{figure*}[ht]
\centering
\includegraphics[scale=0.4]{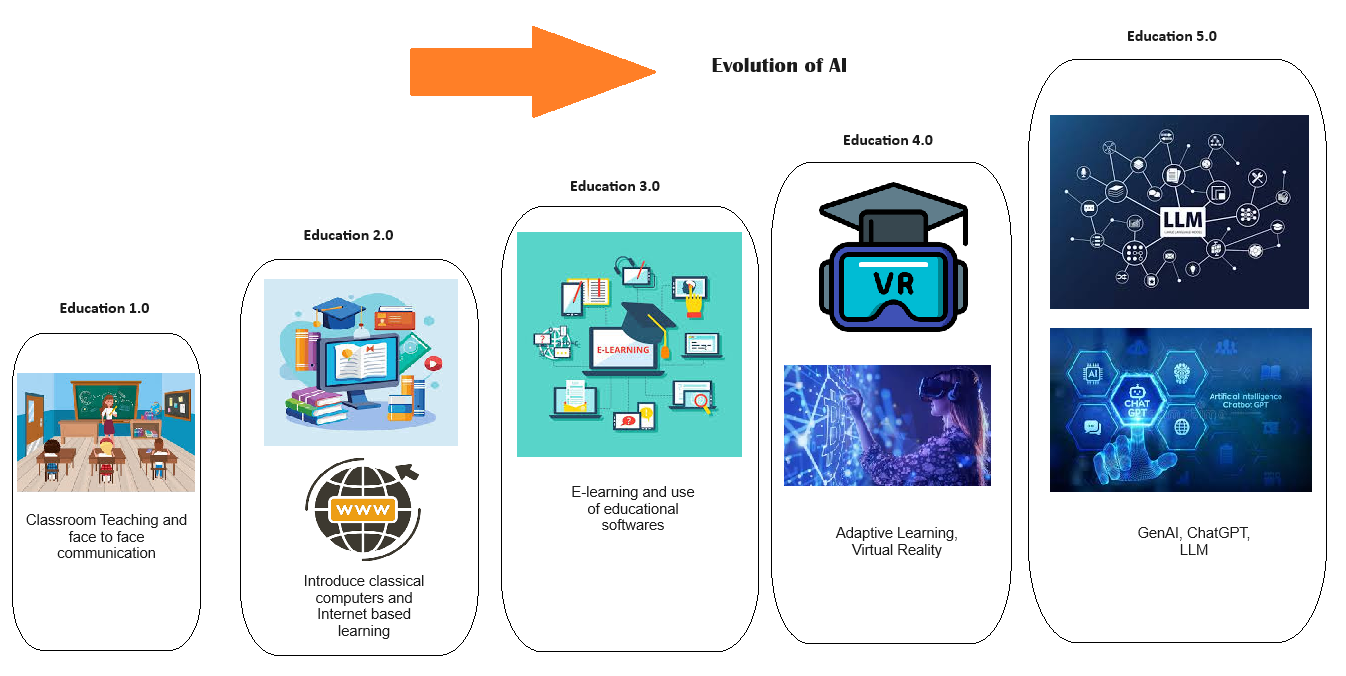}
\caption{Evolution of Education 1.0 to Education 5.0 in the context of AI}
\label{fig1}
\end{figure*}

{\huge\textbf {T}}HE evolution from traditional teacher-centered instruction to a dynamic, technologically-driven, student-centric learning environment is symbolized by the evolution of education from version 1.0 to version 5.0. Technology was not extensively used in education 1.0, which was centred mostly on rote memorisation and standardized curricula. Global networking and collaborative learning were highlighted in Education 3.0, while digital tools were introduced in Education 2.0 as technology progressed. Personalised learning and artificial intelligence (AI) rose to prominence in education 4.0, encouraging creativity and lifetime learning. To create a more adaptable, moral, and learner-centred educational system, Education 5.0 now incorporates AI and holistic approaches, emphasising wellbeing, interdisciplinary studies, and human-AI collaboration \cite{1,10,11}.\\
Artificial Intelligence (AI) is a game-changer in Education 5.0, transforming research, higher education, and the teaching-learning process. When AI is used in education, human-AI interaction is encouraged, which improves individualized learning experiences. AI-powered platforms may customize educational materials to each student's needs, adjust to their learning style, and give immediate feedback. This technology facilitates the transition to a learner-centric approach that emphasizes holistic development, which includes emotional intelligence, creativity, and mental health. Artificial Intelligence (AI) in higher education enables multidisciplinary study by providing tools for solving complicated problems and using virtual and augmented reality (VR/AR) to simulate real-world settings \cite{26}. Significant effects are also seen in research, as AI speeds up data analysis, automates tedious procedures, and creates new opportunities for educational research by facilitating large-scale data collecting and analysis that supports pedagogical advances \cite{19}. Incorporating AI into Education 5.0 fosters global connectivity, sustainability, and ethical thinking in addition to academic success, equipping students to navigate and participate in a world that is changing quickly \cite{1}. It is impossible to overstate the value of AI in research and education. Artificial intelligence (AI)-powered tools and applications have started to change the education system and increase research output. AI can also assist scientists and educators in navigating the increasingly complicated terrain of their respective professions by optimising a variety of research and educational processes, which could ultimately result in ground-breaking discoveries and enhanced learning results \cite{2}. The evolution of education 1.0 to education 5.0 in the context of AI is depicted in Fig \ref{fig1} and summarized in Table \ref{tab1}.

\begin{figure*}[ht]
\centering
\includegraphics[scale=0.9]{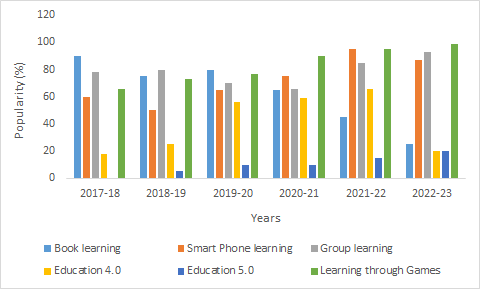}
\caption{Transition in Learning Methodologies over the Past Years}
\label{fig2}
\end{figure*}

Fig \ref{fig2} represents the evolution of learning methodologies in the traditional education system over the past decade. This study has a set of research questions to be answered. They have been listed below.

\begin{enumerate}
\item How does the academic ecosystem integrate GenAI into the process of teaching and learning? 
\item How can the goals of the stakeholders be achieved and the adoption of GenAI in the higher academic setup measured and communicated \cite{60}?
\item What are the various challenges of GenAI in the modern education system? 
\item What are the future directions for enhancing the impact of GenAI in education? 
\end{enumerate}

\begin{figure}[hbt!]
\centering
\captionsetup{justification=centering}
\includegraphics[scale=0.25]{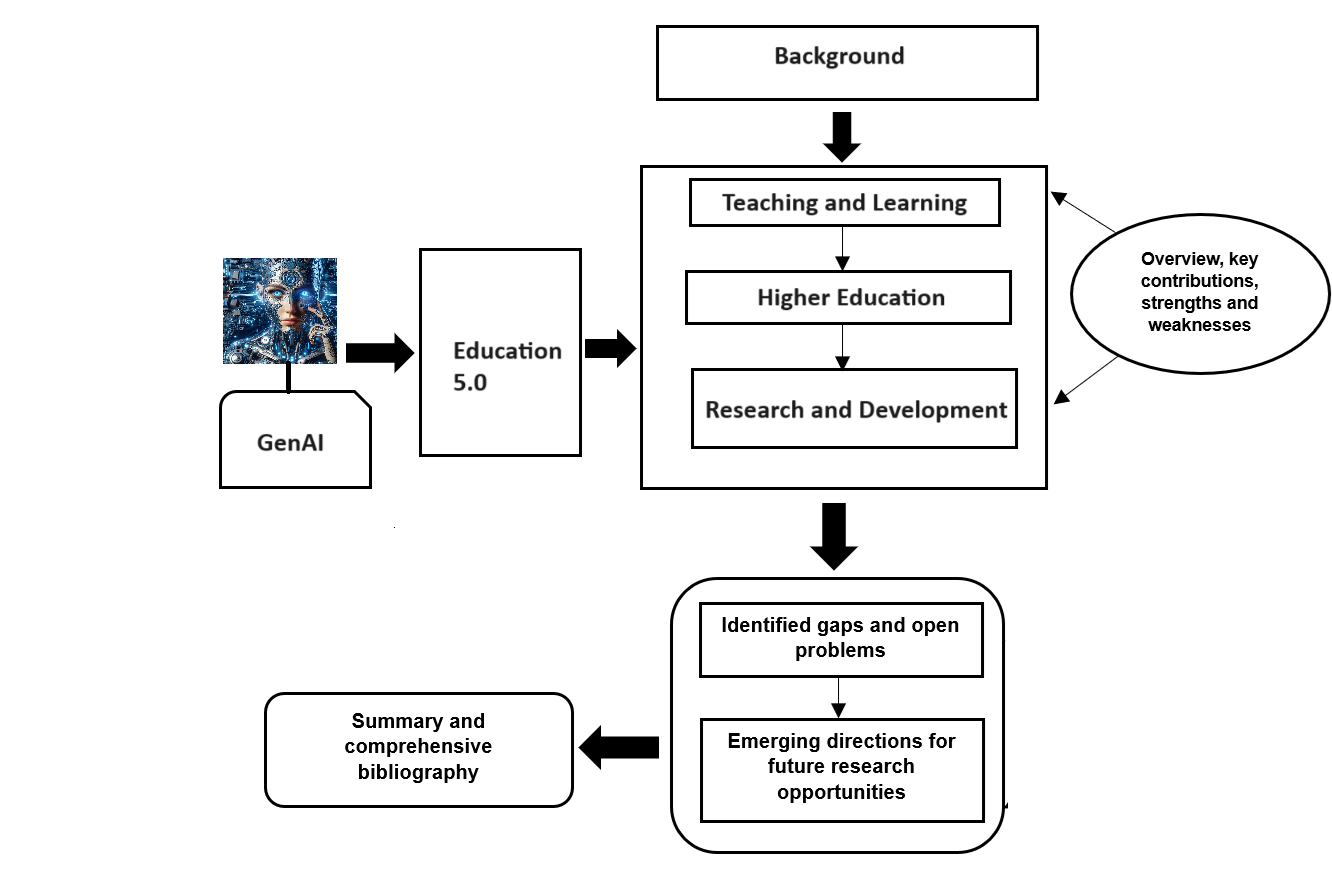}
\caption{Conceptual framework of the study}
\label{figconceptual}
\end{figure}

\begin{table*}[hbt!]
\centering
\caption{Transition from Education 1.0 to Education 5.0 in the context of AI}
\begin{tabular}{p{2cm} p{2cm} p{4cm} p{4cm} } 
 \hline
\\Education Phase &Year &Characteristics &Role of AI\\\\
 \hline
\\ Education 1.0 &Before 1990s  & Teacher centered board learning, Standardized testing and minimal technology use    &Absence of AI\\
 \hline
\\Education 2.0 &1990s-2000s  & Introduction of classical machines and Internet, Collaborative learning, basic online resources   &Basic digital tools, no AI \\
 \hline
\\Education 3.0 &2000s-2010s  & Personalized learning paths, Hybrid learning (in-person + online), Increased use of educational software    &Early AI applications for customization\\
 \hline
\\Education 4.0 &2010s-2020s  & Adaptive machine learning, Real-time data analytics, Experiential learning (Virtual reality, simulations)    &Advanced AI for real-time adaptation and analytics\\
 \hline
\\Education 5.0 &2020s-Present  & Smart teaching and learning, Lifelong, inclusive, and accessible education, AI-driven autonomous learning platforms   &AI deeply integrated into personalized learning, generative learning, large language model (LLM), explainable AI \\
\hline
\end{tabular}
\label{tab1}
\end{table*}
This study seeks to clarify the revolutionary potential of GenAI in education through a detailed examination of the prospects and difficulties as well as a comprehensive analysis of the literature. Three main issues with integrating GenAI into education are outlined in this study. These concerns include the need for,
\begin{itemize} 
\item An overview of the evolution from Education 1.0 to Education 5.0, along with the role of Generative AI in shaping today’s modern and smart education.
\item A clear vision and direction regarding the use of GenAI in education 5.0 through a review of a large number of interesting literature; 
\item An examination of the challenges and gaps associated with the application of GenAI in research, teaching, and learning; 
\item An investigation of the potential applications and future directions of GenAI in the field of modern education.
\end{itemize}
The study’s conceptual framework is shown in Fig. \ref{figconceptual}.
\subsection{Paper Organization}
This study's outline is as follows. The history of generative AI, as well as its practical models and tools appropriate for the educational 5.0 system, are covered in Section II. Section III provides a brief review of the existing literature on the adoption of GenAI in education at different levels worldwide. The gaps and challenges posed by GenAI in education 5.0 are discussed in Section IV. Future directions for researching GenAI in the field of education are also covered. In the end, the study's conclusion is presented in Section V.

\section{BACKGROUND}
This section discusses the fundamental aspects of Education 5.0 and the generative AI tools that can be utilized to enhance the education system \cite{86}.
\subsection{Education 5.0}
The term "education 5.0" describes a new stage in education where cutting-edge technologies and flexible teaching approaches are combined to improve student outcomes. Artificial intelligence (AI), blockchain, the Internet of Things (IoT), and virtual and augmented reality are examples of digital technology. Furthermore, rather than emphasizing only memorization, Education 5.0 emphasizes the development of 21st-century abilities such as creativity, critical thinking, and problem-solving and incorporates immersive learning experiences into the classroom through the use of augmented and mixed reality applications. It is unlikely to be prepared for Education 5.0, however, since it has not fully utilized the potential of Education 4.0, Value-based education, research-based learning, project-based learning, experiential learning, accounting for student aspirations, student flexibilities, curriculum design, teaching-learning-evaluation processes, and outcome-based education are among the most crucial components of Education 5.0 \cite{10}. In order to obtain the maximum benefits, Education 5.0 should be implemented in the following crucial areas \cite{11},
\begin{itemize} 
\item Professional development through concentrated study.
\item A refined and integrated personalised learning concept.
\item Utilising original thought to solve issues.
\item Fostering a culture of value-based learning.
\end{itemize}
The foundation of Education 5.0 is made up of six main pillars: lifelong learning, coherent and relevant curriculum, innovative delivery and assessment, personalised learning, transformative learning, and the incorporation of cutting-edge technology like VR and AI. Additionally, by fusing traditional learning with digital innovations and real-world problem-solving, Education 5.0 focuses on fostering critical thinking, creativity, and flexibility in students, preparing them for a world that is changing quickly \cite{11}. The teacher's function in Education 5.0 shifts from that of a typical instructor to that of a co-learner, resource specialist, mentor, and facilitator. Teachers assist students to explore and make sense of a plethora of information resources by guiding them through personalised learning paths. By promoting critical thinking, creativity, and problem-solving abilities, they motivate students to take an active role in their own educational process. Additionally, educators use cutting-edge technologies, such as artificial intelligence (AI) and digital tools, to improve student learning while maintaining education's inclusivity, flexibility, and relevance to each student's requirements \cite{10,11,94}. 
\subsection{Generative Artificial Intelligence (GenAI)}
Several tools and techniques in Generative AI significantly impact the enhancement of the modern-day educational system, often referred to as the smart educational system \cite{63}. Some of the key technologies are briefly discussed here \cite{58}.\\\\
\textbf{1. GenAI}: One of the most groundbreaking areas in AI is generative AI, which involves systems that can create new and original content. Generative models are capable of producing data that closely matches the input that they are trained on, such as text, images, sounds, or even complex designs, in contrast to typical AI models that concentrate on prediction or classification \cite{5}. There are several types of generative models useful in everyday education, and four of the most widely used GenAI models are discussed. An AI model known as a Generative Adversarial Network (GAN) is made up of two neural networks that operate against one another: the discriminator and the generator. The discriminator determines whether the data is real (from the training set) or fake (from the generator), whereas the generator generates fake data (like images) from random noise. Both of these networks are trained simultaneously: the discriminator sharpens its ability to discern between actual and fake input, and the generator gets better by attempting to trick it. This adversarial process keeps going until the generator generates data that can be mistaken for actual data, leading to the creation of realistic content. One kind of generative model is a variational autoencoder (VAE), which creates new data by learning to represent existing data in a lower-dimensional latent space. There are two primary parts to it: the encoder and the decoder \cite{75}. The encoder extracts the salient characteristics from a variety of potential representations by compressing the input data into a probabilistic latent space. After that, the decoder uses these latent variables to reconstruct the data, making sure that the result is comparable to the original input. Diffusion models are generative frameworks that reverse the process of gradually adding noise to data during training by gradually denoising a sample from a basic noise distribution. The procedure is divided into two primary stages: the forward phase involves gradually adding noise to the input data, and the reverse process involves training a neural network to gradually remove the noise and recover the original data. Deep learning architectures like the transformer model are mainly employed for sequence-to-sequence tasks like text generation and language translation. It can capture long-range relationships without the use of recurrent networks because it uses self-attention mechanisms to assess the relative relevance of various input sequence segments. Multi-head self-attention and feedforward neural network layers include both the encoder and decoder of the transformer \cite{75}.\\
\textbf{2. Large Language Model (LLM)}: A powerful artificial intelligence system that has been trained on a large amount of text to understand, produce, and manipulate human language is known as a Large Language Model (LLM). These models can execute tasks like translation, summarisation, text generation, and question answering with amazing accuracy because they use deep learning techniques, especially neural networks with multiple layers, to identify patterns in language. LLMs are built to do a variety of linguistic tasks and are always improving in understanding and producing natural language. The de facto standard for implementing LLMs in this setting is the use of transformers, specifically generative pretrained transformers (GPT). GPT builds a generative model that duplicates and captures a phrase's structure using copious amounts of textual data. Consequently, LLMs have the ability to process and generate text outputs that resemble those of humans, which opens up a wide range of educational applications \cite{21,24}. \\
\textbf{3. ChatGPT:} A strong language model built on the GPT (Generative Pre-trained Transformer) architecture is the ChatGPT model, created by OpenAI. By processing big datasets and learning linguistic patterns, it excels at comprehending and producing text that is similar to that of a human. ChatGPT is a flexible tool that may be used to have discussions, provide answers to queries, create content, and help with a range of activities in many disciplines. It is a frequently utilised tool in both personal and professional contexts due to its capacity to comprehend the context and give logical, contextually relevant solutions \cite{3,20,27,31,33,47,48,50,53,59,75,77,78,90}. ChatGPT is not the only thing that offers GenAI today. ChagGPT is merely an example of what can be done. There are far more options, which means there are more applications \cite{43}. \\
\textbf{4. Prompting:} The simplest and most common method for changing how general artificial intelligence behaves is to prompt. A set of instructions in plain language outlining appropriate tutoring behaviours, such as this one: "Start by presenting yourself to the student as their AI-Tutor who is happy to help them with any questions," is all that the EdTech designer needs to do \cite{38}. One question at a time. Ask them first what topics they would like to learn more about. Await the reply. That being said, there are certain drawbacks to the prompting strategy. Clearly defining in plain words what excellent teaching behaviours include is necessary. This includes describing all potential rule exceptions, what should be avoided and when, what should be done and when, etc, \cite{18}. In AI, there are three primary forms of prompting: Zero-shot prompting is asking an AI to complete an action and depending just on the prompt itself, without giving any previous examples. Few-shot prompting helps the AI better understand the task by giving it a few samples to influence its response. By encouraging the AI to create intermediate stages or reasoning processes, chain-of-thought prompting produces outputs that are more precise and sophisticated. These prompting strategies aid in improving the AI's output quality and adjusting its performance to particular tasks.

\section{GenAI in Education 5.0}
This section provides a brief overview of the impact of generative AI on the modern education system, based on existing literature. In addition, Table \ref{tab2} summarizes how different GenAI models, tools, and techniques are being applied to enhance various aspects of the education system.
\begin{table*}[hbt!]
\centering
\caption{Outlining Generative AI (GenAI) models suitable for and applied in education 5.0}
\begin{tabular}{p{2cm} p{4cm} p{8cm} } 
 \hline
\\Category &GenAI models/tools &Impact in Education\\\\
 \hline
\\ Content Creation &ChatGPT, Text-to-Text Transfer Transformer (T5), Bidirectional Encoder Representations from Transformers (BERT)  &Generation of personalized lessons and quizzes, Creation of instructional materials tailored to individual needs \cite{34} \\
 \hline
\\Tutoring and Mentoring &GPT-based chatbots  &Real-time assistance, Answering questions, Providing explanations and study guidance \cite{33,93}   \\
 \hline
\\Assessment and Feedback &Automated Grading Tools (e.g., OpenAI’s Codex)  & Automated grading of assignments, Instant feedback on student performance, Customized study plans \cite{24,42}  \\
 \hline
\\Language Learning    &Transformer-based Language Models (e.g., BERT)  & Interactive language exercises, Personalized language learning paths \cite{36,91} \\
 \hline
\\Adaptive Learning Systems  &Adaptive Learning Platforms (e.g., DreamBox, Knewton)  & Adjusting content and difficulty based on student progress, Ensuring personalized and effective learning experiences \cite{65}   \\
\hline
\\Simulation and Virtual Labs    &AI-driven Simulations (e.g., Labster)  & Immersive learning environments, Virtual experiments and labs, Risk-free digital space for practical learning \cite{47}   \\
 \hline
\\Collaborative Learning    &AI-enhanced Collaboration Tools (e.g., Google Classroom with AI features)  & Real-time communication and collaboration, Project management, Peer-to-peer learning facilitation \cite{66} \\
 \hline
\\Accessibility and Inclusivity      &AI for Accessibility (e.g., automatic subtitles, text-to-speech)  & Creation of accessible learning materials, Subtitles, translations, and content tailored for diverse learning needs \cite{41}  \\
 \hline
\\Educational Research and Analytics     &Data Analytics Tools (e.g., AI-driven educational analytics platforms)    & Analysis of learning patterns, Insights into student engagement, Curriculum design and effectiveness evaluation \cite{19,20,29}   \\
\hline
\end{tabular}
\label{tab2}
\end{table*}

\subsection{Generative AI in Teaching and Learning}  
ChatGPT has the potential to enhance both instruction and learning. Students can get individualised coaching and feedback from ChatGPT based on their unique learning requirements and advancement. In order to free up teachers' time to concentrate on other facets of teaching, ChatGPT can be trained to grade student essays in English or art instruction \cite{91}. High school students' writings may be accurately graded by a generative model (ChatGPT) with a correlation of 0.86 with human grades after it was trained on a dataset of essays that had been graded by humans \cite{4}. A work introduces, LearnLM-instructor \cite{18}, a newly developed text-based Gen AI instructor that builds upon Gemini 1.0 and has been further refined for one-on-one conversational tutoring. LearnLM-Tutor uses large language models (LLMs) to adjust to the unique demands and learning preferences of each student, delivering customised instructional help. It is an adaptable tool for educators and learners alike, as it can provide personalised explanations, respond to enquiries, and provide interactive tasks. LearnLM-Tutor seeks to promote more efficient and interesting learning processes by utilising its in-depth knowledge of educational context and material. This will assist students in grasping difficult topics and enhancing their academic achievement \cite{18}. The professional development of teachers can be aided by ChatGPT. By evaluating real-time data from classrooms, AI agents can offer teachers feedback and recommendations on how to improve their instruction (such as how to ask better questions). ChatGPT can help teachers grow professionally by offering fresh teaching concepts including self-regulated learning tasks and activities and learning design methodologies \cite{20}. The outputs of generative models can be tailored to specific circumstances and stimuli. This makes it possible to automatically create customised learning materials based on the requirements, interests, and speed of each student. For a particular student who is having trouble with a math idea, an AI tutor may, for instance, provide on-demand, personalised explanations, examples, and practice problems. In comparison with a standardised curriculum, this kind of personalisation could greatly increase engagement and results \cite{30}. Generative AI (GenAI) enhances geographical and environmental education by creating dynamic simulations and interactive visualizations that help students understand complex ecosystems and geographic processes. GenAI, for example, may create intricate models of the effects of climate change or replicate natural catastrophe situations, providing students with a more in-depth and engaging educational experience that enhances their understanding of environmental concerns \cite{36}. Teachers can overcome long-standing obstacles, encourage critical thinking, and support students' intellectual independence and open-mindedness by incorporating GenAI into English language instruction. According to a study, GenAI has the ability to support inclusive education by offering customised learning opportunities that take into account students' different learning styles and cultural backgrounds. This study highlights five key themes from the interviews, focus groups, and reflective essays. These themes show how GenAI tools, like ChatGPT, may help English language teachers tackle their unique set of teaching issues. Personalised learning experiences, overcoming technology barriers, addressing ideological influences, combating isolation from global trends, and accessible learning materials are some of these themes. \cite{91}.
\subsection{Generative AI in Higher Education}
Higher education systems are dynamic ecosystems that support knowledge co-creation and transfer through a variety of educational approaches, in addition to research. The process of acquiring knowledge involves a complex interaction between teaching and learning, with faculty members serving as both knowledge providers and promoters of an environment that fosters critical thinking, problem-solving, and the development of skills applicable to both current and future careers \cite{68,74}. A work \cite{12} that examines how generative AI explores its potential for educational assessment in higher education, incorporating systems thinking. The shift from a paper-and-pencil format to digital delivery of the same traditional testing items that is, multiple-choice, fill-in-the-blank, and true-or-false question items is a typical critique of the traditional digital assessment. GenAI tools have the ability to help overcome this constraint by creating and delivering complicated assessment tasks for performance evaluations, automated grading and improving the responsiveness and contextualisation of assessment items so that students can use improved simulations and graphics to illustrate their learning. The capacity of AI to provide rapid feedback, personalised and adaptable learning, and learning assignments that are customised based on students' performance levels and abilities is one apparent benefit of utilising GenAI for assessment. Keynote Alta, Realizeit, CogBooks, and ALEKS are a few well-known adaptive learning resources \cite{12}. GenAI's ability to improve students' learning experiences by producing incredibly creative outputs in response to user requests is a crucial application in higher education. ChatGPT and other text-to-text AI generators benefit students especially those who are not native English speakers by providing them with feedback on their writing and idea generation. Meanwhile, text-to-image AI generators, such as Stable Diffusion and DALL-E, are useful resources for imparting technical and creative knowledge in disciplines like design and the arts \cite{16}. According to a study \cite{20}, GenAI has improved management platform performance dramatically by doing jobs, performing routines, and boosting security. To increase productivity, the management team may utilise ChatGPT for (i) creating draft emails and proposals and (ii) data analysis and report generation. By presenting their situations, management teams can ask ChatGPT to suggest actions or choices. Using GenAI technologies, such as ChatGPT, to prioritise learning over assessment results in higher education offers a revolutionary way to teach. Instructors can provide students more than just typical assessment measures by using ChatGPT to provide them personalised, real-time feedback and explanations. This method places more emphasis on ongoing education and comprehension than it does on gauging students' success alone through tests and grades. ChatGPT can help pinpoint areas in which students might require more assistance, recommend other resources, and encourage a more in-depth interaction with the material. This change from a model that is solely assessment-focused to one that emphasises interactive learning and continuous improvement contributes to the development of a more vibrant and encouraging learning environment \cite{28}. According to a study, the integration of generative artificial intelligence tools with an instructional design matrix is essential for creating virtual classrooms for massive open online courses (MOOCs). Personalised and enriched learning experiences can be created and delivered by educators by using these technologies in conjunction with an instructional design matrix.  They offer creative approaches to pique students' interest, modify the material, and encourage individualised learning. Using the 4PADAFE instructional design matrix improves the coherence and efficacy of learning activities even more \cite{34}. Emerging forms of digital literacy include prompt engineering. Because prompt engineering optimises the way AI tools engage with students, it has a significant impact on education and results in more productive learning environments. Teachers can instruct AI models like ChatGPT to provide customised explanations for difficult ideas, such as step-by-step arithmetic solutions or in-depth historical event analysis, by creating unique prompts. Furthermore, intelligent algorithms that are trained with well-designed prompts can produce practice tasks that are tailored to each student's degree of competency in language learning \cite{35}. Instructors frequently invest a significant amount of time and energy in creating lesson plans, creating instructional activities, and assessing the learning objectives of their students. Teachers can automate these tedious processes and concentrate more on developing their teaching strategies and quality control by utilising generative artificial intelligence technology. Additionally, generative AI is capable of analysing learning data and student feedback, which helps teachers pinpoint areas for development and make timely modifications to their methods. By giving teachers individualised, data-driven strategies to improve instruction, student experiences, and overall academic results, the incorporation of generative artificial intelligence into student educational administration offers encouraging prospects \cite{57}. According to a study \cite{64}, including GenAI in a master's-level course on instructional design improves students' familiarity with the technology and their comprehension of its moral ramifications. With an increasing knowledge of the ethical concerns and limitations of GenAI, the majority of students are at the beginning stages of engagement \cite{95}. Themes like better teaching techniques, personal development, and the real-world difficulties of appropriately incorporating GenAI are highlighted in course reflections.
\subsection{Generative AI in Research \& Development}
\par
\begin{figure*}[hbt!]
\centering
\includegraphics[scale=0.3]{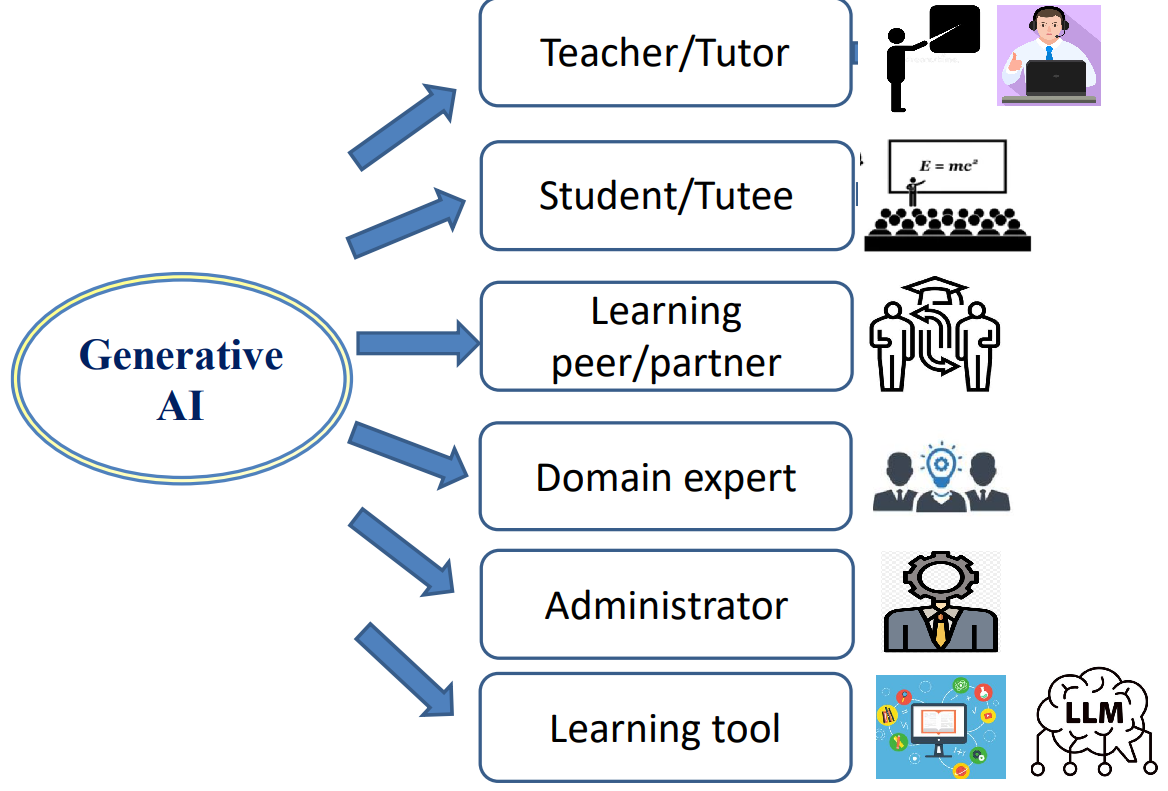}
\caption{Roles of GenAI in Education}
\label{fig3}
\end{figure*}

 Generative AI (GenAI) is enabling creative solutions through sophisticated data-driven models, revolutionizing research and development in the engineering, medical, and agricultural domains. Engineering uses GenAI models such as Variational Autoencoders (VAEs) and Generative Adversarial Networks (GANs) to optimise designs and build complicated structures, like lightweight components in automotive and aerospace engineering, improving efficiency and performance \cite{98}. GenAI has a big impact on drug discovery in the medical industry and R\&D centers. Models anticipate molecular interactions, create new compounds, and help with diagnosis by evaluating patient data and medical pictures to provide tailored therapy \cite{56,77,99}. By evaluating environmental data and genetic information, GenAI is utilised in agriculture to optimise crop yields, create robust plant types, and enhance pest control techniques, resulting in sustainable agricultural practices \cite{100}. These developments highlight the wide range of applications for GenAI in R\&D, since they are stimulating creativity, enhancing decision-making, and promoting sustainability in these vital fields. Fig. \ref{fig3} depicts the roles of GenAI in different sections of an education system \cite{22,76}. Large Language Models (LLMs) are revolutionizing research and development by offering sophisticated features for collaboration, content creation, and data analysis \cite{21}. One noteworthy example is the application of OpenAI's GPT-3 in biomedical research, where it was used to evaluate enormous volumes of scientific literature and produce insights for the development of new drugs. The model was used by researchers to compile results, spot patterns, and even recommend possible medication options based on previous studies. This application demonstrated how LLMs can improve productivity and innovation in complicated disciplines by expediting the research process and facilitating interdisciplinary collaboration. However, in order to maximize their efficacy in research contexts, issues like guaranteeing data quality and combating biases continue to be crucial \cite{24}. In research and development, prompt engineering is essential for maximizing the performance of Large Language Models (LLMs) by improving the input queries to produce more accurate and pertinent results \cite{18}. One such example that demonstrates this is the application of prompt engineering in a project that aims to enhance clinical decision support systems' natural language processing (NLP). In order to assist an LLM in producing medical suggestions based on patient data, researchers created exact prompts \cite{38}. 

\section{Challenges and Future Directions}
This section discusses the fundamental challenges of GenAI in education 5.0 and also explores new future directions in this field.
\subsection{Challenges}
\begin{itemize} 
\item \textit{AI on Social Interaction and Collaborative Learning in Education:} In the learning process, social interaction and collaboration among students, such as group study and peer games, are crucial for engaging with others and gaining diverse perspectives. Students who rely excessively on ChatGPT for answers instead of participating in in-depth discussions with peers and teachers may miss out on acquiring knowledge and developing critical thinking skills through meaningful interactions. The increased use of AI models like ChatGPT poses a risk of diminished face-to-face interaction, potentially hindering social learning \cite{6}.
\item \textit{Risks of Overreliance on ChatGPT:} While ChatGPT might appear to be an effective "teacher" due to its rapid responses, it can create the illusion that students are making swift progress or mastering content quickly. It leads to overconfidence for the students. However, this can lead to an inefficient learning process that exceeds the cognitive load capacity and fails to support deeper understanding \cite{7}.
\item \textit{Enhancing GenAI Learning Models by Integrating Memory, Evaluation, and Human-Centric Dimensions:} The current GenAI model of learning requires two key improvements: first, the integration of long-term memory, and second, the ability to evaluate its outputs and consolidate the knowledge gained from each interaction. More importantly, it lacks the sensory and affective dimensions that are essential to the roles of both teacher and learner. Humans care about one another and about becoming successful learners in addition to acting as behavioural and cognitive agents \cite{8}.
\item \textit{Challenges of Privacy, Data Security, and Academic Integrity:} Privacy, data security, and ethical regulations during assessments using generative AI models are significant challenges in today's context \cite{12,44}. The excessive use of ChatGPT or generative AI in education and research raises challenges related to plagiarism, as it can blur the lines between original work and AI-generated content, making it difficult to distinguish and ensure academic integrity \cite{13,79}.
\item \textit{Challenges for Second Language Learners and Risks of Bias in GenAI-Assisted Writing:} Creating compelling prompts for GenAI tools might be difficult for second language learners since it requires a degree of language skill that they may not have. Furthermore, if students rely too much on GenAI, their real attempts to improve their writing abilities may be compromised. Furthermore, if there are components of bias, inaccuracy, or harm in the underlying training data, these models may produce information that is biased, inaccurate, or harmful \cite{16}.
\item \textit{Risks of Offensive AI-Generated Imagery and Challenges in Detecting AI-Generated Content:} Artificial intelligence-generated imagery has the potential to contain offensive content, such as explicit or nude photos, and can also be utilised maliciously to produce deepfakes. It also becomes difficult to confirm whether a piece of writing is actually the author's original work because AI-generated content frequently escapes existing plagiarism detection techniques \cite{16}.
\item \textit{Job Displacement Fears Among Students Due to the Rise of Generative AI:} The concern that ChatGPT and other GenAI technologies might replace jobs is widespread among students across various fields. As GenAI continues to transform the workplace, there is a growing apprehension that certain careers students are training for may become obsolete \cite{17}.
\item \textit{Impact of AI on Student-Teacher Relationships and Academic Satisfaction:} In academic institutions, there is concern that the pervasive use of AI could impact the student-teacher relationship, potentially leading to student dissatisfaction and a loss of respect for educators \cite{18}.
\item \textit{Addressing Cultural Limitations in Multilingual LLM-Based Education Chatbots:} Multilingual LLM-based education chatbots should target rural families whose children are born in rural areas and go to public schools. Because this type of chatbot is not from a wide cultural background, some of its viewpoints and ideas may be constrained \cite{42}.
\item \textit{Importance of Emotional Connection and Holistic Education in the Age of AI:} Interactions between humans and AI can yield knowledge, but they cannot take the place of the fundamentals of sympathetic and emotional connection. It is imperative to prioritise education and cultivate well-rounded individuals while utilising artificial intelligence (AI) to disseminate knowledge. In order to develop holistic and well-rounded people, efforts should be focused on improving pupils' character, intelligence, physical fitness, aesthetics, and practical talents \cite{57}.
\end{itemize}

\subsection{Future Directions}
\begin{itemize} 
\item \textit{Leveraging Reinforcement Learning:} To overcome the challenge of lack of social interactive learning \cite{6}, reinforcement learning algorithms enable AI models to learn from the feedback and experiences of other agents within a social context. For instance, an AI agent can improve its gameplay by observing the strategies and performance of other agents, then adjusting its own strategy based on the outcomes of these interactions. 
\item \textit{Enhance Competitiveness and Flexibility:} Universities are making an effort to equip their scholars and students for a future in which artificial intelligence (AI) tools such as ChatGPT will be essential in the workplace. This will help graduates stay competitive and flexible in an ever-changing job market \cite{9}.
\item \textit{Roles of Educators and Developers in Ensuring Equitable Access to Generative AI Tools:} What parts do educators and developers play in developing and preserving students' access to generative AI tools? What effects might be seen on the fairness of the educational experience for students across various groups, as well as on their short-term and long-term academic and non-academic performance, and how can equitable and sustainable access be ensured \cite{13}?
\item \textit{Exploring Digital Literacy and Curriculum Development:} The advent of GenAI has prompted scholars to explore the meaning of digital literacy in the age of automation and algorithms. There is a greater likelihood that AI, media, data, computational, and algorithm literacy will have a substantial impact on higher education. There are different meanings for these literacies and AI literacy that are not exclusive. Future research on the advancement of these literacies and their interrelationships needs to be done \cite{19}. New curriculum frameworks and benchmarks for interdisciplinary or multidisciplinary teaching or programs are proposed and evaluated by future research \cite{29}. 
\item \textit{Navigating the Responsible Use of Generative AI:} Students must decide when and how to apply GenAI. While AI can help with low-human contact activities like data analysis, project planning, and coding, using it to write historical essays or research requires a thorough comprehension of context. Verifying facts is necessary, and interacting with others becomes essential. It is essential to know how to use the appropriate equipment for the right jobs \cite{25}.
\item \textit{Evaluating the Impact of Generative AI on Student Wellbeing and Mental Health:} Future studies should evaluate how generative AI affects students' stress, anxiety, and mental health, especially in light of concerns about excessive screen usage and a decline in in-person contact. It is essential to make sure that wellbeing is a priority in the design of GenAI technologies \cite{30}.
\item \textit{Integrating Generative AI into Curriculum and Teaching Methods:} It is also recommended that more research be done on how GenAI might be included in curriculums and teaching methods like entrepreneurship education and inquiry-based learning. This involves assessing the usefulness of GenAI in various educational contexts, such as language acquisition and STEM education, and its function in formative evaluations \cite{41}.
\item \textit{Expanding Research on Adapting Generative AI:} More study is still required to determine how to adapt GenAI to a range of educational contexts and other types of data. To expand the breadth and depth of inquiry, investigations could incorporate materials such as theses and books in addition to databases and geographic regions. In order to fully capture the changing perspectives and long-term effects of GenAI on instructional strategies and learning objectives, this methodology should incorporate longitudinal studies \cite{41}.
\item \textit{Training Educators to Integrate Generative AI in Classrooms:} 90\% of students said that, despite AI's promise for education, their teachers do not encourage them to use the technology as a teaching aid in their classrooms. This suggests that there is a big disconnect between students' interest in and readiness to use AI and teachers' encouragement or assistance. To fully utilise them in the classroom, educators will need to be trained and educated in these new technologies in the future \cite{51}. To effectively navigate and incorporate generative AI with open educational resources (OER) and open educational practices (OEP), educators and stakeholders need to possess the requisite skills and knowledge. This calls for a particular emphasis on professional development and training \cite{61}. 
\item \textit{Addressing Student Misuse of ChatGPT:} Many students claimed that their primary use of ChatGPT was for non-academic objectives; they gave various reasons for this, including breaking academic policies, relying too much on technology, producing assignments that lacked creativity, and possible security threats. Instead of using ChatGPT for group or in-class projects, students mostly use it for assignments. Students reported that they had not been instructed on how to use ChatGPT in a secure and efficient manner. As a result, students must receive in-depth instruction on ChatGPT and 'prompting' in future \cite{84}.
\end{itemize}

\section{CONCLUSION}
This study presents the evolution of education 1.0 to education 5.0 in the context of AI. It provides an in-depth discussion, grounded in existing literature, on the role of generative AI tools and technologies in modern educational society, particularly in classroom teaching and learning, higher education, and research \& development. Finally, the study identifies existing knowledge gaps and challenges, offering future directions aimed at bridging these gaps and inspiring further research in the field.

\end{document}